\documentclass[twocolumn,showpacs,preprintnumbers,amsmath,amssymb,aps]{revtex4-1}
\usepackage{graphicx}% Include figure files
\usepackage{bm}% bold math
\usepackage{dcolumn} % Align table columns on decimal point
\usepackage{subfigure}
\usepackage{color}
\usepackage{epsfig}

\begin{document}

\title{Simulations of metastable states near the apex of a force microscope tip interacting with an ionic crystalline surface}
\author{B.~Ittermann$^{1}$, R.~Hoffmann-Vogel$^{1*}$ and A.~Baratoff$^{2}$}
\affiliation{$^{1}$ Physikalisches Institut and DFG-Center for Functional
Nanostructures (CFN), Karlsruhe Institute of Technology Campus South, D-76128
Karlsruhe, Germany\\
$^{2}$ National Center of Competence in Research (NCCR)
on Nanoscale Science, Institute of Physics, University of Basel,
Klingelbergstrasse 82, CH-4056 Basel, Switzerland\\
$^*$email: r.hoffmann@kit.edu\\
}
\date{\today}
\begin{abstract}
%Loosely bound atoms on rocksalt type scanning force microscopy (SFM) tips
Atoms or pairs of ions picked up by probe tips used in dynamic
force microscopy (DFM) can be strongly displaced and even hop
discontinuously upon approach to the sample surface. The energy
barriers for some of those hops are of the right order of
magnitude to explain the rise in energy dissipation commonly
observed in DFM measurements at room temperature. The systematic
computations reported here can explain the infrequent jumps and
very low average energy dissipation observed low temperature in a
previous DFM study on a KBr(001) sample.  Close to the surface we
indeed find new states separated by small energy barriers which
account for those phenomena. These energy barriers strongly depend
on details of the atomic arrangement in the vicinity of the tip
apex.
\end{abstract}
%\pacs{68.37.Ps 34.20.Cf 68.43.De}
%Comment for preprint

\maketitle

\section{Introduction}
Dynamic force microscopy (DFM) has developed into a valuable tool
not only for surface characterization of non-conducting samples,
but also for controlled modification at the atomic level. This
has become possible mainly thanks to sensitive measurement modes
where the tip is oscillated with a constant amplitude in the nm
range at a resonance frequency of the force sensor \cite{moritab}.
Atomic-scale precision is then achieved if the tip apex stays or
periodically comes to distances at which short-range
site-selective forces act, thereby causing a measurable frequency
shift.  Atom manipulation experiments in that mode have inspired
computations of changes in the potential landscape induced by the
tip apex and of resulting fingerprints in measurable quantities on
semiconductor surfaces
\cite{pizzagalli03,dieska05,oyabu06,custance09p1}, as well as on
ionic crystal surfaces \cite{trevethan06b,canova11p1}. The average
dissipation of energy stored in the cantilever oscillation also
exhibits atomic-scale contrast, even on defect-free surfaces, and
its magnitude indicates that it mainly originates from hysteretic
hopping of atoms between two or more stable positions
\cite{sasaki00p1,kantorovich04}. Sudden but infrequent contrast
changes, typically more pronounced in dissipation images, have
been attributed to long-lived rearrangements of the tip apex
\cite{bennewitz00}.

Three different causes of dissipation induced by hysteretic
tip-sample interactions must be considered: hopping on the sample,
hopping between tip and surface, and hopping on the tip. Hopping
on the sample is usually not expected because diffusion or other
rearrangements on clean flat terraces of low-index surfaces
usually involve rather high energy barriers, except for some
reconstructed surfaces which exhibit bistable configurations
\cite{kantorovich06p1}. The presence of long-lived localized
defects can be excluded by taking high-resolution images.
%although it is known for example for molecules in
However, in scanning tunneling microscopy studies \cite{dunphy93},
as well as for DFM on insulating surfaces, mobile adsorbates can
merely lead to blurry or streaky images and to additional noise in
a certain temperature range thus causing blurred or streaky images
when the scan and hopping rates roughly match \cite{watkins07p1}.
%Energy dissipation due to
Hopping between tip and sample \cite{sasaki00p1,kantorovich04},
that can in the extreme case even lead to atomic chain formation
and breaking in some oscillation cycles \cite{kawai11p1},
manifests itself indirectly via the average energy dissipation.
% or directly via the the above-mentioned sudden contrast changes.
%provides additional information about the hysteretic tip-sample interaction.
However, in order to unambiguously interpret measured results, the
third scenario, hopping on the tip, must be excluded or else taken
into account \cite{hoffmannr07p1,ghasemi08p1}. This is in general
difficult because the structure and chemical composition of the
tip are unknown. For many commonly studied crystals (Si, KBr,
NaCl), there are indications that sample material is picked up by
the tip owing to intentional or accidental contact prior to or
during DFM measurements. Using large-scale simulations several
groups have characterized force microscope tips and derived
construction principles for realistic model tips from comparisons
with experiments \cite{oyabu06,canova11p1,kawai11p1}.

\section{Model tip} \label{TIP}
\begin{figure}
        \begin{center}
        \includegraphics[width=0.8\linewidth,angle=0,clip]{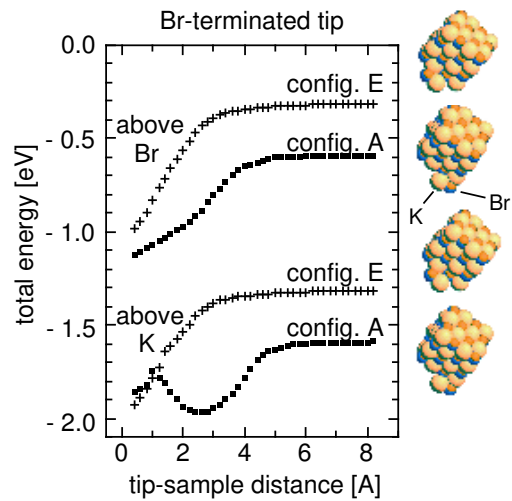}
        \end{center}
    \caption{(color online) Binding energy with respect to an
arbitrary offset chosen for visual clarity as a function of
decreasing nominal tip-sample distance for the Br$^-$ terminated
tip in configurations A and E illustrated on the right. In both
cases the rigid top half of the tip is approached along the same
path such that the protruding Br$^-$ in configuration A is
initially above a Br$^-$ or a K$^+$ surface ion. The data
calculated above K$^+$ is offset by -1 eV for clarity.}
    \label{fig1}
\end{figure}

Here, we study possible low-energy configurations of an overall
neutral KBr tip supporting two additional ions, and its
interaction with a KBr (001) surface by means of extensive
computations. The employed code, based on an atomistic shell
model, was developed for simulations of DFM on ionic crystals
\cite{kantorovich00,bennewitz00} and validated in previous studies
\cite{hoffmannr04p1,ruschmeier08p1}. Such a model tip may
represent the end of a nominal silicon tip typically used in force
microscopy experiments decorated by sample material. More
precisely, our model tip consists of
%The tip was represented by
a K$^+$-terminated cubic cluster of 64 K and Br ions exposing
stable \{001\}-facets oriented such that the (111) direction is
perpendicular to the sample surface. One K$^+$ and one Br$^-$ ion
are added near one of the $<$100$>$ edges meeting at the tip apex,
%of the cubic tip in order to provide a model
as illustrated in Fig. \ref{fig1} right. The initial
configuration of these two additional ions is chosen in accordance
with simulations and a previous experimental study of diffusion on
surfaces of rocksalt-type crystals \cite{shluger95p1}.  This model
is well-suited to for studying rearrangements of the simplest
moiety likely to be picked upon gentle contact with the sample.
Alternatively, the supported KBr dimer may be the remnant of a
broken chain of ions formed during tip retraction
\cite{lantz06p1,kawai11p1}. The assumed tip configuration is
probably more likely than alternative ones involving other kinds
of defects which produce appreciable force hysteresis and energy
dissipation in the case of chemically similar NaCl model tips
\cite{canova11p1}.  The sample was represented by a slab
containing 6 layers of 10 $\times$ 10 ions each. More details
about the simulation procedure can be found in previous
publications \cite{hoffmannr04p1,ruschmeier08p1}.

First, the properties of the decorated tip alone were studied.
%Five different
One stable and four metastable configurations labelled A to E were
found by constrained minimization while shifting the additional
Br$^-$ ion parallel to the edge (projected reaction coordinate
$q$) and letting its two orthogonal coordinates and those of all
ions in the bottom half of the cube relax. The corresponding
profiles labelled $z=$ inf are shown in Fig. \ref{fig2}.
In configurations A and E, the additional Br$^-$ and K$^+$ ions
are essentially located along the cube edge, and the positions of
the Br$^-$ ion differ between both configurations by approximately
one bulk lattice constant. The lowest total energy was found for
the (Br$^-$-terminated) configuration A, because the electrostatic
field outside the cube is enhanced in the vicinity of the
low-coordinated edge especially around the apex. For the same
reason the energy barrier to reach configuration A starting from E
is much lower than for the opposite process. Configurations B and
D arise when the additional Br$^-$ ion is located above a bridge
site on one facet adjacent to the cube edge. Between $q=2.1$ {\AA}
and $q=2.9$ {\AA} the initial Br$^-$ ion dips into the cubic
cluster while a nearby bromine ion emerges from the cluster to
form configuration C. Ion exchange processes analogous to that
just described were also found in previous simulations of an MgO
dimer diffusing on the MgO (001) surface\cite{henkelman05p1}.
Similar configurations were found when all Br$^-$ ions were
replaced by K$^+$ ions and vice-versa.

To test the thermal stability of the different tip configurations,
molecular dynamics simulations were performed at $T=200$, $300$
and $500$ K for the K$^+$ and the Br$^-$ terminated tip in
configuration A. Below 500 K, no hopping was observed over the
relatively short simulation time. At $T=500$ K the former tip
showed a transition from the E to the A configuration nicely
visualized in a movie \cite{footnote0}. In this transition the
K$^+$ ion did not, however, jump directly to the final position.
Instead, it moved to the position of a nearby K$^+$ ion in a
configuration similar to that  called C in the constrained
minimizations mentioned earlier. This exchanged K$^+$ ion then
moved to the position of the K$^+$ ion in the A configuration.
Such exchange processes compete against pivoting around the dimer
partner as in diffusion on (001) surfaces
\cite{shluger95p1,henkelman05p1} and may even be favored in the
present lower coordination situation.

\section{Approach to the surface - new configurations}
In a second step, the interaction of the tip with the sample
surface and possible hysteretic processes were studied as a
function of the nominal tip-sample distance defined as the
separation of the foremost ions when ignoring relaxation. Ions in
the top half of the tip cluster and in two boundary layers of the
sample slab were frozen, while the rest were allowed to relax. The
rigid part of the tip was incrementally approached perpendicular
to the surface such that the protruding Br$^-$ in configuration A
was facing a particular surface ion. The results were compared to
configuration E at the same positions of the rigid tip body (Fig.
\ref{fig1}). If the tip is approached above an ion of the
same charge, the energy difference between configurations A and E
becomes smaller, but never vanishes. In contrast, if the tip is
approached above an oppositely charged ion, the energy first drops
faster then the energy difference decreases and vanishes for the
K$^+$-terminated tip, and even becomes negative for the
Br$^-$-terminated tip below a critical tip-sample distance of
about 1 \AA. One therefore expects the Br$^-$-terminated tip to
change from configuration A to configuration E below that distance
if the energy barrier between the two states can be overcome by
thermal fluctuations.

\begin{figure}
        \begin{center}
        \includegraphics[width=\linewidth,angle=0,clip]{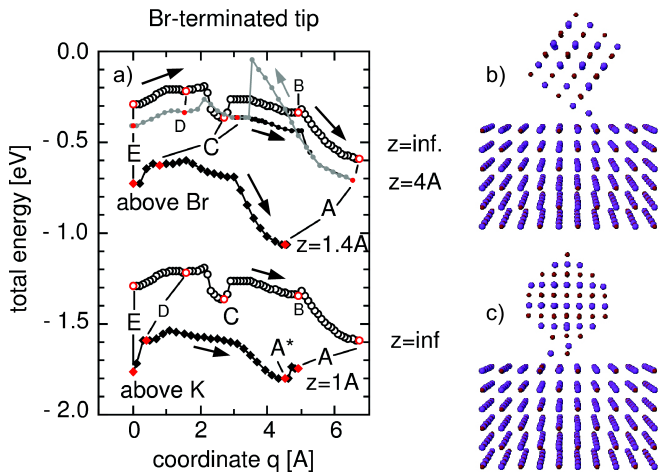}
        \end{center}
    \caption{(color online) a) Energy profiles calculated for the
%K$^+$-terminated decorated
initially protruding Br$^-$ tip ion above equally and oppositely
charged surface ions at a few nominal tip-sample distances. The
coordinate $q$ of the moved Br$^-$ points along the cube edge away
from its apex. The data above K$^+$ has been offset by -1 eV for
clarity. All black curves represent constrained minimizations at
$q$-increments of 0.1 \AA. The red points refer to subsequent full
minimizations. b) and c) Side views on the new configuration A*,
viewed along the $x$- and along the $y$-directions, respectively.}
    \label{fig2}
\end{figure}

In order to investigate changes in the energy landscape
%between the two states
induced by the tip approach, we performed constrained
minimizations like those discussed in Section \ref{TIP} at a few
tip-sample distances. Figure \ref{fig2} shows that the
number and character of the metastable configurations changes
significantly. While at relatively large distances five
%stable or metastable
configurations analogous to A, B, C, D and E are observed, only
three remain at close tip-sample distances above the Br$^-$
surface ion. Lateral hysteresis is also observed, e. g. for $z=1$
\AA, indicating the presence of inequivalent energy barriers along
paths starting from configurations A and E. A new configuration
(A*), even lower in energy than configuration A, appears above the
oppositely charged K$^+$ surface ion for $z=1$ \AA. In
configuration A* the additional K$^+$ and Br$^-$ ions of the
molecule are bound to both tip and surface. This configuration
arises when the body of the cubic tip pushes the
%additional molecule
added dimer aside, until the dimer ions approximately bind to ions
of opposite charge on the surface as well as on a tip facet, as
illustrated in another set of movies \cite{footnote1}.

\section{Energy barriers for configurational changes}

Our study originally focused on hopping between A and E
configurations, because simulations of diffusion on rocksalt type
(001) surfaces identified this process as the most probable
\cite{shluger95p1}. The presence of intermediate metastable
configurations implies that direct hopping between A and E is less
probable than hopping via the intermediate states B, C and D, due
to the reduced intervening energy barriers which are the relevant
ones for thermal activation. The highest energy barrier between
any of those configurations represents the bottleneck of the
process and therefore determines the effective hopping rate. The
bottleneck barrier for the transition from E to A ranges from 80
to 175 meV while the bottleneck barrier for the opposite
transition from A to E ranges from 260 to 660 meV for all of the
constrained energy profiles studied \cite{footnote3}.
%For the jumps between intermediate states smaller barriers are found.
These values should be compared to the thermal energy at room
temperature (25 meV) or at low temperatures (8 - 40 K
\cite{hoffmannr07p1}, i. e. $0.7 - 3.4$ meV) depending on the
experiment to be considered.  Two limiting cases are of particular
interest \cite{hoffmannr07p1,ghasemi08p1}: if the hopping rate is
low enough (one jump every $0.1 - 10$ s), individual atomic jumps
can be experimentally observed. If, in contrast, the hopping rate
exceeds the cantilever oscillation frequency ($\sim 100-200$ kHz)
the individual states involved are averaged over in a dynamic
force measurement, but the energy dissipation due to hops into
lower energy configurations becomes appreciable. Our results imply
that, for most potential energy landscapes so far considered, the
tip configuration would be rapidly flipping at room temperature on
time scales faster than the cantilever oscillation, but that tip
changes due to hops would be frozen out at low temperatures. The
main reason is that in configuration A, the
%hopping ion is located on the low-coordinated cube-edge which
protruding Br$^-$ ion is subjected to the positive electrostatic
potential localized around the apex of the cube \cite{sushko99}.
This lowers its binding energy and thus raises the
%overall
bottleneck energy barrier between A and E. Therefore, hopping
between A and E can only account for
%weak energy dissipation
infrequent jumps at room temperature, but not for the single jumps
observed at low temperatures \cite{hoffmannr07p1}.

\begin{table}[t]
\begin{tabular}{c|c|c|c|c}
surface&tip-sample&transition&energy barrier&bottle-\\
site&distance [\AA]&path& [meV]&neck \\\hline\hline
&&A$\rightarrow$B&272&$\times$\\
&&B$\rightarrow$C&78&\\
\raisebox{1.5ex}[-1.5ex]{-}&\raisebox{1.5ex}[-1.5ex]{$\infty$}&C$\rightarrow$D&176&\\
&&D$\rightarrow$E&7&\\ \hline
&&A$\rightarrow$C&662&$\times$\\
&4.0&C$\rightarrow$D&94&\\
Br&&D$\rightarrow$E&4&\\\cline{2-5}
&&A$\rightarrow$C&461&$\times$\\
&\raisebox{1.5ex}[-1.5ex]{1.4}&C$\rightarrow$E&3&\\\hline
&&A$\rightarrow$A*&10&\\
K&1.0&A*$\rightarrow$D&264&$\times$\\
&&D$\rightarrow$E&2&\\\hline\hline
&&E$\rightarrow$D&88&\\
&&D$\rightarrow$C&25&\\
\raisebox{1.5ex}[-1.5ex]{-}&\raisebox{1.5ex}[-1.5ex]{$\infty$}&C$\rightarrow$B&103&$\times$\\
&&B$\rightarrow$A&23&\\ \hline
&&E$\rightarrow$D&83&$\times$\\
&4.0&D$\rightarrow$C&67&\\
Br&&C$\rightarrow$A&0.1&\\\cline{2-5}
&&E$\rightarrow$C&107&$\times$\\
&\raisebox{1.5ex}[-1.5ex]{1.4}&C$\rightarrow$A&22&\\\hline
&&E$\rightarrow$D&174&$\times$\\
K&1.0&D$\rightarrow$A*&51&\\
&&A*$\rightarrow$A&64&\\\hline
\end{tabular}
\caption{Summary of all the energy barriers obtained in
constrained minimizations above different surface sites.}
\label{table1}
\end{table}

As an alternative model tip one may consider a KBr cubic cluster
with the additional K$^+$ and Br$^-$ ions placed in positions of
higher coordination such as a facet. However, with this premise it
is more difficult to realize an atomically sharp tip, as required
for
%atomic resolution imaging
lattice-resolved images without appreciable distortions
\cite{oja05p1} observed experimentally.

\section{Low energy barriers}

The jumps observed at low temperatures might be explained by the
occurrence of new states similar to A*. Indeed, the energy barrier
between A and A* being only 64 meV at a tip-sample distance of 1
\AA, individual jumps could be observed at temperatures between 25
and 30 K.

%neuer Text
\begin{figure}
        \begin{center}
        \includegraphics[width=\linewidth,angle=0,clip]{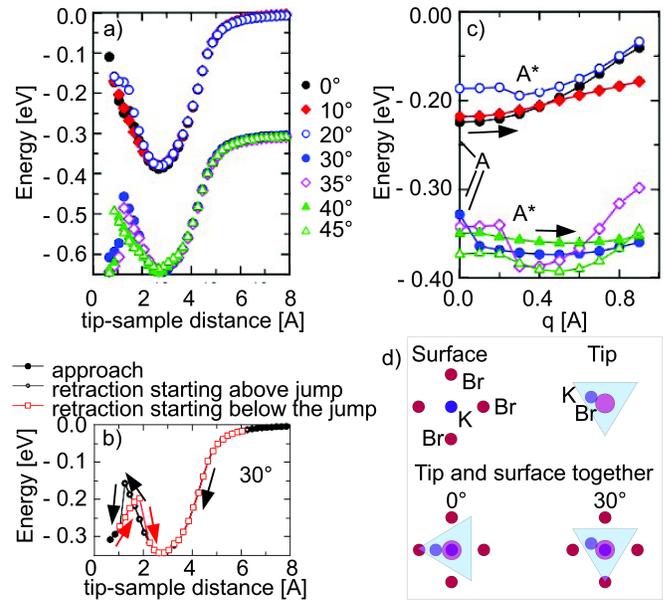}
        \end{center}
    \caption{(color online) Interaction energy as a function of distance
for different polar angles of the Br$^-$-terminated tip with
respect to the [001] axis passing through a surface K$^+$ site. a)
Between 0 and 20$^{\circ}$ no jump occurs, while between 30 and
45$^{\circ}$ (curves offset by -0.3 eV for clarity) hysteresis is
observed in the investigated distance range. b)
Approach-retraction hysteresis observed for 30$^{\circ}$. c)
Results of constrained minimizations for orientations where no
jump was observed and (curves offset by -0.3 eV for clarity) for
those where hysteresis was observed. d) Schematic drawing
illustrating two inequivalent orientations of the tip cluster with
the initially on-edge adsorbed dimer with respect to a surface
unit cell viewed from above.}
    \label{fig3}
\end{figure}

In order to investigate the latter possibility in more detail, we
calculated the total energy as a function of tip-sample distance
for inequivalent rotations of the
%K$^+$-terminated decorated
Br$^-$-terminated tip around the $z$-axis perpendicular to the
sample surface passing through a surface K$^+$ site, starting from
alignment of the added KBr dimer with one of the $<$100$>$ axes of
the sample at 0$^{\circ}$. In this notation our previous
calculations correspond to 30$^{\circ}$. Due to the symmetries of
the surface and the model tip, full information is contained for
polar angles between 0$^{\circ}$ and 45$^{\circ}$. At distances
larger than about 2.0 \AA, the data are qualitatively and
quantitatively similar for all rotation angles (Fig.
\ref{fig3} a) and b)). However, at smaller distances
significant differences occur. Between 30$^{\circ}$ and
45$^{\circ}$, a jump in the energy arises at an angle-dependent
critical distance below maximum attraction. At smaller distances,
the interaction energy decreases again upon approach. This
behavior can be identified with the state A* discussed above, as
evident in movies of the approach \cite{footnote1}. In contrast,
between 0$^{\circ}$ and 20$^{\circ}$, the energy continuously
increases further upon approach in the same distance regime. The
foremost tip ion then remains in a deformed state A roughly under
the tip apex, while for angles between 30$^{\circ}$ and
45$^{\circ}$, the foremost tip atom jumps towards a next neighbor
surface ion of opposite sign while the dimer becomes aligned with
a surface $<$100$>$ direction, as well as with a tip facet in
configuration A*. For 20$^{\circ}$ the energy versus distance
curve is deformed around the critical distance, an indication of
the proximity to an additional state, but the tip remains in the
deformed state A.

\begin{table}[t]
\begin{tabular}{c|c|c|c|c}
rotation&critical&critical&hysteresis&energy\\
angle&distance of&distance of&loop&barrier\\
&approach&retraction&area&A*$\rightarrow$A\\
$^{\circ}$&[\AA]&[\AA]&[meV]&[meV]\\ \hline
20&-&-&-&19\\
30&1.3 & 2.1 &200&91\\
35&1.3 & 2.1 &170&93\\
40&1.1 & 2.7 &190&23\\
45&0.9 & 1.9 &90&40
\end{tabular}
\caption{Critical distances, hysteresis loop areas and energy
barriers for different tip orientations. To obtain the data, the
tip was first approached to the surface up to a distance of 1.3
{\AA}, then the
%ions were moved to
dimer was forced into state A* by the constraint, the system was
then fully relaxed
%in state A*
and finally the tip was retracted. For 20$^{\circ}$ a
configuration similar to A* is only reached under the constraint
but, upon retraction from this configuration, the two added ions
remain on the surface, so that critical distances and hysteresis
%cannot be calculated
are not observed.} \label{table2}
\end{table}

We finally studied the stability of configuration A*. When the tip
is retracted starting from distances larger than critical, the
initial values of the energy, force and atomic positions are
smoothly recovered. Otherwise hysteresis is observed until
eventually another jump restores the initial energy, force and
atomic positions at a larger tip-sample critical distance. In Fig.
\ref{fig3} c) an example for 30$^{\circ}$ is shown which
is also further documented in a movie of the simulated retraction
\cite{footnote2}. Since our calculations are done at zero
temperature, the observed hysteresis implies that the energy
barrier between A and A* vanishes at the critical distance of
approach and that energy is gained by jumping to A* at closer
tip-sample distances. Similarly, upon retraction energy is gained
by jumping back to state A while the reverse energy barrier
vanishes at the critical distance of retraction. Between the two
critical points a finite energy barrier exists between the two
states. This variation of the energy landscape as a function of
tip-sample distance corresponds to the scenario proposed by Sasaki
and Tsukada \cite{sasaki00p1} with the modification that the
%jumps
atomic hops themselves need not occur in the $z$ direction between
the tip and the sample but must only be induced by the tip motion
perpendicular to the sample surface.  Results obtained for
different relative orientations of the tip are summarized in table
\ref{table2}.  The energy loss
%caused by the hysteretic jump
in approach-retraction cycle of the tip is equal to the area
enclosed between the two distinct force-distance curves between
both critical distances and amounts to up to 0.2 eV which is in
the range of what is expected from low-temperature experiments
\cite{hoffmannr07p1}.
%for a single jump (table \ref{table2}).

We further characterized the stability of configuration A* by
studying the energy barrier from state A to state A* in
constrained minimizations along the previously defined reaction
coordinate $q$. The tip-sample distance was chosen to be 0.13 nm
because this is the largest distance at which state A* is observed
during approach and in
%the intermediate
a finite distance range upon retraction, and so we expected that
the state could be observed for several tip angles. Indeed,
between 30$^{\circ}$ and 40$^{\circ}$, where A* was observed in
energy vs. distance data, state A* is also observed in constrained
minimizations. For 20$^{\circ}$, although state A* is not observed
during approach, the system can be driven into a similar state
under the constraint. The energy barriers range between 19 and 91
meV depending on the tip orientation. These energy barriers are of
the right order of magnitude for explaining the infrequent jumps
seen in low temperature experiments \cite{hoffmannr07p1}. One
should, however, consider that the chosen reaction coordinate $q$,
was not modified to take into account the tip deformation.
%at close tip-sample distances.
It is therefore possible that even lower energy barriers occur in
other directions, in particular at close tip-sample distances.
Obviously, the precise value of the energy barriers depend on the
details of the atomic arrangement of the tip.

\section{Conclusions}

We conclude that
%moveable ions on rocksalt-covered scanning
ion pairs picked by force microscope tips decorated by sample
material can be strongly displaced upon approach to the sample
surface, in particular at close tip-sample distances where they
become bound to both tip and sample. The
%strong deformation of the tip can result in jumps in the direction
resulting hops have components parallel to the sample surface.
%Such jumps can explain
Some of those hops can account for rapid flipping of the tip
configuration at room temperature. In addition, some metastable
states which occur at close tip-sample distances can be separated
by energy barriers that are low enough to explain infrequent
individual jumps observed at low temperatures. The
%moveable entities
adsorbed dimers have a tendency to align with ions of opposite
charge on the sample surface. Ion exchange processes previously
identified in a study of surface diffusion are preferred
%additionally play a role.
for some of the investigated hops.  Back and forth hops between
metastable configurations result in a hysteretic force as a
function of distance. The energy dissipated by such hops is in the
range of what is expected from low-temperature experiments
\cite{hoffmannr07p1}.
%what is expected experimentally for a single jump.

%\section{Acknowledgements}

Financial support from the Landesstiftung Baden-W\"urttemberg  in
the framework of its excellence program for postdoctoral
researchers, from the European Research Council through the
Starting Grant NANOCONTACTS (No. 239838) and from the NCCR
Nanoscale Science of the Swiss National Science Foundation is
gratefully acknowledged.


\begin{thebibliography}{99}
\bibitem{footnote3}
The results shown in Fig. 2 were
%mainly calculated
obtained in the direction from E to A with the exception of the
path shown in grey
%for the position
above a surface Br$^-$ for $z=4$ \AA. For all other cases the
barriers were assumed to be the same independent of the direction.
\bibitem{footnote2}
For a rotation angle of 20$^{\circ}$, the
%additional atoms of the
dimer ions remain on the surface upon retraction if the tip is
initially approached
%to a distance
closer than 0.35 nm to the surface. The final state with the two
additional ions transferred to the surface has an energy 310 meV
lower than configuration E, because the dimer is now in a higher
coordinated state on the surface compared to its location on a
 tip
edge in configuration E.
\end{thebibliography}
\end{document}